\documentclass[runningheads]{llncs}

\usepackage{cite}
\usepackage{mathtools}
\usepackage{amsmath}

\usepackage{amsmath,amssymb,amsfonts}
\usepackage{algorithm}
\usepackage{algorithmic}

\usepackage{graphicx}
\usepackage{textcomp}
\usepackage{xcolor}

\makeatletter

\floatstyle{ruled}
\newfloat{algorithm}{tbp}{loa}
\providecommand{\algorithmname}{Algorithm}
\floatname{algorithm}{\protect\algorithmname}

\usepackage[caption=false,font=footnotesize]{subfig}
\usepackage{algorithmic}

\floatname{algorithm}{Algorithm}

\makeatother

\begin{document}

\title{Energy-Efficient MIMO Multiuser Systems: Nash Equilibrium Analysis}
\author{Hang Zou \inst{1} \and  Chao Zhang \inst{1} \and  Samson Lasaulce \inst{1} \and  Lucas Saludjian \inst{2} \and Patrick Panciatici \inst{2}}

\authorrunning{H. Zou et al.}

\institute{L2S, CNRS-CentraleSupelec-Univ. Paris Saclay, Gif-sur-Yvette, France \and RTE, France\\
\email{\{hang.zou, chao.zhang, samson.lasaucle\}@centralesupelec.fr}}

\maketitle
\begin{abstract}
In this paper, an energy efficiency (EE) game in a MIMO multiple access
channel (MAC) communication system is considered. The existence and
the uniqueness of the Nash Equilibrium (NE) is affirmed. A bisection
search algorithm is designed to find this unique NE. Despite being  sub-optimal for deploying the $\varepsilon$-approximate
NE of the game when the number of antennas in transmitter is unequal
to receiver's, the policy found by the proposed algorithm is shown
to be more efficient than the classical allocation techniques. Moreover,
compared to the general algorithm based on fractional programming
technique, our proposed algorithm is easier to implement. Simulation
shows that even the policy found by proposed algorithm is not the
NE of the game, the deviation w.r.t. to the exact NE is small and
the resulted policy actually Pareto-dominates the unique NE of the
game at least for 2-user situation.

\end{abstract}

\vspace{-0.1in}
\section{Introduction \label{sec:Introduction}}

With the release of first 5G package, it turns out that the number
of devices in  the upcoming wireless network will increase
tremendously, e.g., Internet of Things (IoT). Consequently, classical
paradigm which merely aims at optimizing the quantitative performance,
e.g., data-rate, bit-error-rate and latency faces extreme difficulty
in many domains in both academic research and industrial application.
Thus the issue of energy-efficient design of the wireless system tends
to be crucial. Different definition of energy efficiency (EE) has been proposed in
recent years in \cite{ng-TWC-2012,he-TOC-2013,vetu-TWC-2015,du-CL-2014}. Amongst which the most popular one is defined as the
total benefit obtained under the unit consumption of energy or power known as global energy efficency (GEE) e.g., in \cite{richter-VTC-2009,belmega-TSP-2011,zapp-TWC-2013,zhang-TVT-2019}. Taken
the bits-per-second type rate function as benefit function, one will
obtain well-known the bits-per-joule energy efficiency.

One of the pioneer works of studying the maximization of EE in Multiple-Input Multiple-Output (MIMO) system is \cite{belmega-TSP-2011}.
In \cite{belmega-TSP-2011}, the optimal precoding scheme is studied and divided into
different cases with different assumptions on the systems. Till now
the optimal precoding matrix for general condition is merely conjectured
and unproved. Hereafter, optimal precoding matrix design for single
user MIMO system is performed for imperfect channel state information (CSI) scenario in \cite{varma-TSP-2013}. Then it
is later widely realized that the problem of EE maximization actually
belongs to the category of fractional programming. Techniques such
as Dinkelbach's algorithm (see \cite{dinkel-MS-1967}) is used to solve EE maximization in \cite{ragha-ICC-2010 , varma-TSP-2013}. These algorithms are generally based on the
idea that the optimal solution can be found by solving a sequence
of convex optimization problems related to the original one. The main
difficulty of EE maximization OP is usually due to the non-convexity
of energy efficiency function. Under some assumption on the benefit
function, the EE function is well-known as being quasi-concave or
even pseudo-concave. However, it is generally difficult to trace the
Nash Equilibrium (NE) of a game where the individual utility function
of player is of EE type. In \cite{zapp-TWC-2013}, it is shown that there always exists
an unique NE for scalar power allocation game in a relay-assisted
MIMO systems due to the standard property of the best response dynamics.
Similar results in MIMO-MAC system will be given latter in the paper.

The contribution of this paper is twofold: 1) we first extend the work in \cite{zapp-TWC-2013} to a more
general situation where each user is allowed to choose its covariance matrix
to maximize its individual EE instead of tuning its scalar power merely.
The existence and uniqueness of the NE is proved under some assumptions. 2) An algorithm is proposed to to find the unique NE of this MIMO-MAC
game. When the number of antennas of transmitter is equal  the one of receiver,
proposed algorithm leads to exact NE. Otherwise it leads to the $\varepsilon$-approximate
NE defined latter in the paper for replacing the exact best response
dynamic by its linear approximation.

The remaining parts of the paper is organized as follows: the 
MIMO-MAC system and the EE game are first presented in Sec. \ref{sec:System_model}. Then some basics
of game theory are given and the existence and the uniqueness of NE
of the EE game is proved in Sec. \ref{sec:Game_theoretic_Analysis}. In Sec. \ref{sec:Algorithm_for_NE}, a basic algorithm
is proposed and an improved bisection search algorithm  is given which yields 
an $\varepsilon$-approximate NE slightly Pareto-dominating the exact NE. The numeric results of proposed
algorithms are compared with classical allocation policy and
analyzed in Sec. \ref{sec:Numeric_Results}.  The paper concludes by several remarkable
conclusions in Sec. \ref{sec:Conclusions}.

Notations: $\left(\cdot\right)^{H}$ and
$\left(\cdot\right)^{\dagger}$ denote matrix transpose and Moore-Penrose inverse respectively. $\mathbf{I}_{N}$
stands for identity matrix of size $N$. $\text{det}\left(\cdot\right)$
and $\textrm{Tr}\left(\cdot\right)$ denote the determinant and the
trace of a matrix respectively. Denote the natural number set inferior
or equal than $N$ as $\left[N\right]\triangleq\left\{ 1,\dots,N\right\} $.
\vspace{-0.1in}

\section{System model\label{sec:System_model}}

Consider a multiple access channel (MAC) with one base station (BS)
and $K$ users (players) to be served. BS is equipped with $N_{r}$
receive antennas and each user terminal is equipped with $N_{t}$
transmit antennas. We assume a block fading channel where the realization
of channel remains a constant during the coherence time of transmission
and randomly generated according to some statistical distributions from
period to period. The received signal at BS is given by:
\vspace{-0.02in}
\begin{equation}
\boldsymbol{y}=\sum_{k=1}^{K}\mathbf{H}_{k}\boldsymbol{x}_{k}+\boldsymbol{z},\label{eq:general_MIMO_system}
\end{equation}
\vspace{-0.02in}
where $\mathbf{H}_{k}\triangleq\left[\mathbf{H}_{k,i,j}\right]_{i,j=1}^{N_{r},N_{t}}\in\mathbb{C}^{N_{r}\times N_{t}}$
is the channel transmit matrix of $k$-th user and $\mathbf{H}_{k,i,j}$
is the channel from $i$-th transmit antenna of $k$-th user to $j$-th
receive antenna at BS which is assumed to be i.i.d. complex Gaussian
distributed according to $\mathcal{CN}\left(0,1\right)$. $\boldsymbol{x}_{k}=\left(x_{k,1},\dots,x_{k,N_{t}}\right)^{T}$
is the transmit symbol of $k$-th user and $\boldsymbol{z}$ is the
noise observed by the receiver with complex Gaussian distribution
$\mathcal{CN}\left(\boldsymbol{0},\sigma^{2}\mathbf{I}_{N_{r}}\right)$.
For the sake of simplicity, we assume that single user decoding is
implemented for each user. Then the capacity the $k$-user can be
achieved is 
\begin{equation}
R_{k}=\log\frac{\det\left(\sigma^{2}\mathbf{I}_{N_{r}}+\sum_{j=1}^{K}\mathbf{H}_{j}\mathbf{Q}_{j}\mathbf{H}_{j}^{H}\right)}{\det\left(\sigma^{2}\mathbf{I}_{N_{r}}+\sum_{j\neq k}^{K}\mathbf{H}_{j}\mathbf{Q}_{j}\mathbf{H}_{j}^{H}\right)},\label{eq:capacity_type_rate}
\end{equation}
where $\mathbf{Q}_{k}=\mathbb{E}\left[\boldsymbol{x}_{k}\boldsymbol{x}_{k}^{H}\right]\in\mathbb{C}^{N_{t}\times N_{t}}$
is the covariance matrix of symbol $\boldsymbol{x}_{k}$ which determines
how power should be allocated for each antenna and $P_{c}>0$ is the
power dissipated in transmitter circuit to operate the devices. It
is reasonable to assume that each user has perfect knowledge about
its own channel, e.g., through downlink pilot training. Therefore
user $k$ is able to perform the singular value decomposition (SVD)
of its own channel $\mathbf{H}_{k}$ and its covariance matrix $\mathbf{Q}_{k}$
as well. The SVD of $\mathbf{H}_{k}$ and $\mathbf{Q}_{k}$ is given
by $\mathbf{H}_{k}=\mathbf{U}_{k}\mathbf{\Lambda}_{k}\mathbf{V}_{k}^{H}$
and $\mathbf{Q}_{k}=\mathbf{W}_{k}\mathbf{P}_{k}\mathbf{W}_{k}^{H}$
respectively. To simplify the problem, we assume that user $k$ always
adapts its covariance matrix to $\mathbf{H}_{k}$, i.e., choosing
$\mathbf{W}_{k}=\mathbf{V}_{k}$. $\mathbf{P}_{k}$ is a diagonal
matrix with $\mathbf{P}_{k}=\textrm{diag}\left(\boldsymbol{p}_{k}\right)=\textrm{diag}\left(p_{k1},\dots,p_{kN_{t}}\right)$ where we use $\textrm{diag}\left( \cdot \right)$ to generate a diagonal matrix from a vector or vice versa.
Thus user $k$'s only legal action is represented by $\boldsymbol{p}_{k}$
or $\mathbf{P}_{k}$ and the action set of $k$-th user is
\begin{equation}
\mathcal{P}_{k}=\left\{ \boldsymbol{p}_{k}\left|\sum_{i=1}^{N_{t}}p_{ki}\leq\overline{P}_{k},\ p_{ki}\geq0\right.\right\} \label{eq:action_set_covariance_continuous}
\end{equation}
where $\overline{P}_{k}$ is power budget of $k$-th user. Through
out the paper, we will use the matrix $\mathbf{P}_{k}$ or its diagonal
$\boldsymbol{p}_{k}$ interchangeably to represent user $k$'s action
depending on the context. Further more, we denote $\boldsymbol{p}=\left(\boldsymbol{p}_{k},\boldsymbol{p}_{-k}\right)$
with $\boldsymbol{p}_{-k}\triangleq\left(\boldsymbol{p}_{1},\dots,\boldsymbol{p}_{k-1},\boldsymbol{p}_{k+1}\dots,\boldsymbol{p}_{K}\right)\in\mathcal{P}_{-k}$
and $\mathcal{P}_{-k}\triangleq\mathcal{P}_{1}\times\dots\times\mathcal{P}_{k-1}\times\mathcal{P}_{k+1}\times\dots\times\mathcal{P}_{K}$.
In this paper energy efficiency defined as the ratio between a benefit
function and the power consumed by producing it can be proven to has
the following expression for user $k$ after some simplifications:
\begin{align}
u_{k}\left(\mathbf{P}_{k},\mathbf{P}_{-k}\right) & =\frac{\log\frac{\text{det}\left(\sigma^{2}\mathbf{I}_{N_{r}}+\sum_{j=1}^{K}\mathbf{U}_{j}\mathbf{\Lambda}_{j}\mathbf{P}_{j}\mathbf{\Lambda}_{j}^{H}\mathbf{U}_{j}^{H}\right)}{\text{det}\left(\sigma^{2}\mathbf{I}_{N_{r}}+\sum_{j\neq k}^{K}\mathbf{U}_{j}\mathbf{\Lambda}_{j}\mathbf{P}_{j}\mathbf{\Lambda}_{j}^{H}\mathbf{U}_{j}^{H}\right)}}{\textrm{Tr}\left(\mathbf{P}_{k}\right)+P_{c}}\label{eq:EE_utility_function}
\end{align}
To this end, the MIMO MAC EE game is thus given by the following strategic
form in triplet:
\begin{equation}
\mathcal{G}=\left(\mathcal{K},\left(\mathcal{P}_{k}\right)_{k\in\mathcal{K}},\left(u_{k}\right)_{k\in\mathcal{K}}\right)\label{eq:MIMO_GEE_games_continuous}
\end{equation}

\section{Game-theoretic Analysis \label{sec:Game_theoretic_Analysis}}
In this section, we will firstly give some basic concepts of any game-theoretic
analysis. The central concept of game-theoretic analysis is Nash Equilibrium
(NE) defined as:
\begin{definition}
For game $\mathcal{G}=\left(\mathcal{K},\left(\mathcal{P}_{k}\right)_{k\in\mathcal{K}},\left(u_{k}\right)_{k\in\mathcal{K}}\right)$,
an action profile $\boldsymbol{p}=\left(\boldsymbol{p}_{k},\boldsymbol{p}_{-k}\right)$
is called a Nash Equilibrium if for $\forall k\in\mathcal{K}$ and
$\forall\boldsymbol{p}'=\left(\boldsymbol{p}'_{k},\boldsymbol{p}_{-k}\right)$:
\vspace{-0.05in}
\begin{equation}
u_{k}\left(\boldsymbol{p}_{k},\boldsymbol{p}_{-k}\right)\geq u_{k}\left(\boldsymbol{p}'_{k},\boldsymbol{p}_{-k}\right)\label{eq:NE_definition}
\end{equation}
\end{definition}
The meaning of  NE is that any unilateral change of action at this
point won't lead to an enhance of individual benefit. Furthermore,
we introduce an important conception in game-theoretic analysis known
as best response dynamics.
\begin{definition}
(Best Response): In a non-cooperative game $\mathcal{G}$, the correspondence
$\mathrm{BR}_{k}\left(\boldsymbol{p}_{-k}\right)$: $\mathcal{P}_{-k}\rightarrow\mathcal{P}_{k}$
s.t.
\vspace{-0.01in}
\begin{equation}
\mathrm{BR}_{k}\left(\boldsymbol{p}_{-k}\right)\triangleq\arg\max_{\boldsymbol{p}_{k}\in\mathcal{P}_{k}}u_{k}\left(\boldsymbol{p}_{k},\boldsymbol{p}_{-k}\right)\label{eq:definition_of_best_responce}
\end{equation}
\vspace{-0.01in}
is called the best response (BR) of player $k\in\mathcal{K}$ given
the action profile of other player $\boldsymbol{p}_{-k}$. From the
definition of best response, one has immediately the following characterization
for NE:
\end{definition}
\begin{proposition}
{[}Nash,1950{]} An action profile $\boldsymbol{p}^{\star}$ is an
NE if and only if : $\forall k\in\mathcal{K}$,\ $\boldsymbol{p}_{k}^{\star}\in\mathrm{BR}_{k}\left(\boldsymbol{p}_{-k}^{\star}\right)$.
\end{proposition}
To identify the NE of game in (\ref{eq:MIMO_GEE_games_continuous}),
the properties of individual utility function should be identified
as first step. We define two critical properties satisfied by the
individual utility function.
\begin{definition}
(Quasi-concavity) Let $\mathcal{X}\in\mathbb{R}^{n}$ be a convex
set, a function $f:\mathcal{X}\rightarrow\mathbb{R}$ is said to be
quasi-concave if 
\begin{equation}
f\left(\lambda\boldsymbol{x}+\left(1-\lambda\right)\boldsymbol{y}\right)\geq\min\left\{ f\left(\boldsymbol{x}\right),f\left(\boldsymbol{y}\right)\right\} \label{eq:quasiconcavity_defintion}
\end{equation}
for any $\boldsymbol{x},\boldsymbol{y}\in\mathcal{X}$ with $\boldsymbol{x}\neq\boldsymbol{y}$
and $0<\lambda<1$.
\end{definition}
\begin{definition}
(Pseudo-concavity) Let $\mathcal{X}\in\mathbb{R}^{n}$ be a convex
set, a function $f:\mathcal{X}\rightarrow\mathbb{R}$ is said to be
quasi-concave if it is differentiable and for any $\boldsymbol{x},\boldsymbol{y}\in\mathcal{X}$,
it holds:
\begin{equation}
f\left(\boldsymbol{y}\right)<f\left(\boldsymbol{x}\right)\implies\nabla f\left(\boldsymbol{y}\right)^{T}\left(\boldsymbol{x}-\boldsymbol{y}\right)>0
\label{eq:pseudoconcavity_definition}
\end{equation}
\end{definition}
With the definition of quasi-concavity and the pseudo-concavity, Prop.
\ref{prop:convexity_of_utility_and_benefits} shows that the individual
utility function does possess these important properties:
\begin{proposition}
\label{prop:convexity_of_utility_and_benefits} $R_{k}$ is a concave
functions w.r.t. $\boldsymbol{p}_{k}$ \textcolor{black}{and $u_{k}$
is a pseudo-concave (quasi-concave) function w.r.t. }$\boldsymbol{p}_{k}$
for $\forall k\in\mathcal{K}$; For any fixed $\boldsymbol{p}_{-k}\in\mathcal{P}_{-k}$
and $p_{kj}$ with $j\neq i$, only one of following statements is
true for all $i\in\left[N_{t}\right]$:

i) $\exists\ p_{ki}^{\star}>0$ s.t. $u_{k}$ is an increasing function
in $\left(0,p_{ki}^{\star}\right)$ and a decreasing function in $\left(p_{ki}^{\star},+\infty\right)$
w.r.t. $p_{ki}$.

ii) $u_{k}$ is a decreasing function in $\left(0,+\infty\right)$
w.r.t. $p_{ki}$.
\end{proposition}
\begin{proof}
\textcolor{black}{It is well-known that $R_{k}$ is a concave function
for $\boldsymbol{p}_{k}$. Then the pseudo-concavity (quasi-concavity)
of $u_{k}$ comes from the fact that it is a ratio of a concave function
and an affine function of $\boldsymbol{p}_{k}$. For more details
of the proof, see \cite{zapp-sur-2015}.  Now we prove the second part of this
proposition.} 
Rewrite the individual utility function as $u_{k}=\frac{R_{k}\text{\ensuremath{\left(\gamma_{k}\right)}}}{\sum_{i=1}^{N_{t}}p_{ki}+P_{c}}$
with $R_{k}\left(\gamma_{k}\right)=\log\left(1+\gamma_{k}\right)$.
Then we can prove that $\frac{\partial^{2}u_{k}}{\partial p_{ki}^{2}}\leq0$
due to the fact that $R_{k}$ is an increasing concave function w.r.t.
$\gamma_{k}$ and $\gamma_{k}$ is a also increasing concave function
w.r.t. $p_{ki}$. However we can't conclude directly of the sign of
$\lim_{p_{ki}\rightarrow+\infty}\frac{\partial u_{k}}{\partial p_{ki}}$.
It can be positive or negative depending on the value of $p_{kj}$
with $j\neq i$. Therefore, if $\lim_{p_{ki}\rightarrow+\infty}\frac{\partial u_{k}}{\partial p_{ki}}\geq0$
then we are in case ii), otherwise we are in case i).
\end{proof}

Before stating the best response dynamics of the game, we define the
following boundary of set $\mathcal{P}_{k}$ indicated by an index
subset $\mathcal{E}\subset \left[N_{t}\right]$ :
\begin{equation}
\mathcal{P}_{k}\left[\mathcal{E}\right]\triangleq\left\{ \boldsymbol{p}_{k}\in\mathcal{P}_{k},p_{ki}=0\ \textrm{for}\ i\in\mathcal{E}\right\} \label{eq:boundary_action_set}
\end{equation}
and the non-negative index set for a given action $\mathbf{P}_{k}$:
\begin{equation}
\mathcal{I}\left(\mathbf{P}_{k}\right)\triangleq\left\{ i\in\left[N_{t}\right]s.t.\ p_{ki}\geq0\right\} \label{eq:non_negative_index_set}
\end{equation}

\begin{proposition}
\label{prop:NE_EE_MIMO_continuous} For any given $\mathbf{P}_{-k}$
and provided that the power budget $\overline{P}_{k}$ is sufficiently
large, denote the unique solution of the following equation as $\mathbf{P}_{k}^{*}$:
\begin{equation}
\textrm{diag}\left(\mathbf{\Lambda}_{k}^{H}\left(\mathbf{\Lambda}_{k}\mathbf{P}_{k}\mathbf{\Lambda}_{k}^{H}+\mathbf{F}_{k}+\sigma^{2}\mathbf{I}_{r}\right)^{-1}\mathbf{\Lambda}_{k}\right)
= u_{k}\left(\mathbf{P}_{k},\mathbf{P}_{-k}\right)\mathbf{I}_{N_{t}}\label{eq:pseudo_best_response}
\end{equation}
 
with $\mathbf{F}_{k}=\sum_{j\neq k}\text{\ensuremath{\mathbf{S}}}_{j}\mathbf{P}_{j}\mathbf{S}_{j}^{H}$
is the interference matrix of $k$-th user with $\mathbf{S}_{j}=\mathbf{U}_{k}^{H}\mathbf{U}_{j}\mathbf{\Lambda}_{j}$.
Then the BR of $\mathbf{P}_{k}$ w.r.t. $\mathbf{P}_{-k}$ is standard
and converges to the unique NE admitted by game (\ref{eq:MIMO_GEE_games_continuous});
The BR is the unique solution of (\ref{eq:pseudo_best_response})
restricted to the boundary of $\mathcal{P}_{k}$ indicated by $\mathcal{I}\left(\mathbf{P}_{k}^{*}\right)$
with $\mathcal{I}\left(\mathbf{P}_{k}^{*}\right)\neq\emptyset$.
\end{proposition}
\begin{proof}
our proof consists of two parts: i) existence of NE; ii) uniqueness
of NE.
i) Existence of NE: it is easy to prove that the action set $\mathcal{P}_{k}$
for each player is compact (closed and bounded), combining the quasi-concavity
of $u_{k}$ claimed in Prop. \ref{prop:convexity_of_utility_and_benefits},
the existence is due to Debreu-Fan-Glicksberg theorem \cite{debreu-NAS-1952}.
Moreover, \textcolor{black}{Prop. \ref{prop:convexity_of_utility_and_benefits}
claims that $u_{k}$ is a pseudo-concave function w.r.t. }$\mathbf{P}_{k}$.\textcolor{black}{{}
Due to the property of pseudo-concave function, the unique stationary
point (points where derivative vanishes) is the global optimizer of
the utility function if the stationary point is in the feasible action
set. We first calculate the stationary point of $u_{k}$ for $\forall k\in\mathcal{K}$
using matrix calculus which leads to }(\ref{eq:pseudo_best_response}).
However, the stationary point might not belong to the feasible action
set $\mathcal{P}_{k}$. Denote $\mathbf{P}_{k}^{*}$
the unique solution of (\ref{eq:pseudo_best_response}) in $\mathbb{R}^{N_{t}}$.
It is easy to prove that for given $\mathbf{P}_{-k}$, $p_{ki}^{*}$ is a decreasing function  w.r.t.  $\forall p_{kj}$
with $j\neq i$ by contradiction.
Due to this monotonicity of the BR and knowing that the feasible action
set $\mathcal{P}_{k}$ is a polyhedron, BR must be on the boundary
of $\mathcal{P}_{k}$ except $\mathbf{0}_{N_{t}\times N_{t}}$ defined
as (\ref{eq:boundary_action_set})  which corresponds to the index set $\mathcal{I}\left(\boldsymbol{p}_{k}^{*}\right)\neq\emptyset$
, which completes the proof for existence.

ii) Now we would like to prove that the BR converges to a point which
is the unique NE of the game. \textcolor{black}{We will achieve that
by showing that the best response is a standard function }\footnote{\textcolor{black}{The generalized inequality for matrix defined here is
referred to its diagonal and takes the non-negative orthant as the
underlying cone.}}\textcolor{black}{, i.e.,}

\textcolor{black}{1) Positivity: $\forall\ \mathbf{P}_{-k}\succcurlyeq0$,
$\textrm{BR}_{k}\left(\mathbf{P}_{-k}\right)\succcurlyeq0$;}

\textcolor{black}{2) Monotonicity: if $\mathbf{P}^{'}{}_{-k}\succcurlyeq\mathbf{P}_{-k}$,
then $\textrm{BR}_{k}\left(\mathbf{P}^{'}{}_{-k}\right)\succcurlyeq\textrm{BR}_{k}\left(\mathbf{P}_{-k}\right)$;}

\textcolor{black}{3) Scalability: $\textrm{BR}_{k}\left(\alpha\mathbf{P}_{-k}\right)\prec\alpha\textrm{BR}_{k}\left(\mathbf{P}_{-k}\right)$
for any $\alpha>1$.}

Positivity is obviously observed in its form given by Prop. \ref{prop:NE_EE_MIMO_continuous}.
The proof for\textcolor{black}{{} }monotinicity and scalability is similar
to \cite{zapp-TWC-2013}. The strict proof is omitted due to the limit of space.
\end{proof}
\vspace{-0.2in}
\section{Algorithm for finding NE\label{sec:Algorithm_for_NE}}

Prop. \ref{prop:NE_EE_MIMO_continuous} actually provides an approach
for us to find the NE of the game (\ref{eq:MIMO_GEE_games_continuous}).
One can easily deduce an iterative equation according to (\ref{eq:pseudo_best_response}):
\begin{equation}
  \textrm{diag}\left(\mathbf{\Lambda}_{k}^{H}\left(\mathbf{\Lambda}_{k}\mathbf{P}_{k}^{\left(t\right)}\mathbf{\Lambda}_{k}^{H}+\mathbf{F}_{k}^{\left(t-1\right)}+\sigma^{2}\mathbf{I}_{r}\right)^{-1}\mathbf{\Lambda}_{k}\right)
 = u_{k}\left(\mathbf{P}_{k}^{\left(t-1\right)},\mathbf{P}_{-k}^{\left(t-1\right)}\right)\mathbf{I}_{N_{t}}\label{eq:iterative_equation_solve_eq}
\end{equation}
However, due to Prop. \ref{prop:NE_EE_MIMO_continuous}, this stationary
point might not be in the feasible action set. One can design the
following basic algorithm to find NE of the game (\ref{eq:MIMO_GEE_games_continuous})
based on Prop. \ref{prop:NE_EE_MIMO_continuous} summarized in alg.
\ref{alg:basic_algorithm_for_NE}.

\begin{algorithm}[ht]
\begin{algorithmic}

\STATE\ Initialization: $\mathbf{P}_{k}^{\left(0\right)}=\frac{1}{N_{t}}\mathbf{I}_{N_{t}},\forall k$.
Choose $T$ and $\epsilon$

\STATE\ $\mathbf{For}$ $t=1$ to $T$, $\mathbf{do}$

\STATE\ \ \ $\mathbf{For}$ $k=1$ to $K$, $\mathbf{do}$

\STATE\ \ \ \ \ \ \ Compute $\mathbf{P}_{k}^{\left(t\right)}$
using (\ref{eq:iterative_equation_solve_eq})

\STATE\ \ \ \ \ \ \ $\mathbf{If}$ $\mathcal{I}\left(\mathbf{P}_{k}^{\left(t\right)}\right)\neq\left[N_{t}\right]$

\STATE\ \ \ \ \ \ \ \ \ \ \ Compute $\mathbf{P}_{k}^{\left(t\right)}$
using (\ref{eq:iterative_equation_solve_eq}) restricted to $\mathcal{I}\left(\mathbf{P}_{k}^{\left(t\right)}\right)$

\STATE\ \ \ \ \ \ \ $\mathbf{End}$ $\mathbf{If}$

\STATE\ \ \ $\mathbf{End}$ $\mathbf{For}$

\STATE\ \ \ $\mathbf{If}$ $\sum_{k}\left\Vert \mathbf{P}_{k}^{\left(t\right)}-\mathbf{P}_{k}^{\left(t-1\right)}\right\Vert <\epsilon$

\STATE\ \ \ \ \ \ \ $\mathbf{Break}$

\STATE\ \ \ $\mathbf{End}$ $\mathbf{If}$

\STATE\ $\mathbf{End}$ $\mathbf{For}$

\STATE\ Output: $\mathbf{P}_{k}^{\textrm{NE}}=\mathbf{P}_{k}^{\left(t\right)}$
for $\forall k$.

\end{algorithmic}

\caption{Basic Algorithm for finding NE of MIMO-MAC EE game\label{alg:basic_algorithm_for_NE}}
\end{algorithm}
Nevertheless, alg. \ref{alg:basic_algorithm_for_NE} is not satisfatory
way to find the NE of the game. More precisely, to find the BR for
given $\mathbf{P}_{-k}$, one actually need to solve an optimization
problem. However, if $h=U\left(\mathbf{P}_{-k}\right)=\max_{\mathbf{P}_{k}\in\mathcal{P}_{k}}u_{k}\left(\mathbf{P}_{k},\mathbf{P}_{-k}\right)$
is known as \emph{a priori} information, (\ref{eq:iterative_equation_solve_eq})
can be transformed into following equation which is relatively easy
to be solved compared to (\ref{eq:iterative_equation_solve_eq}) :
\begin{equation}
\textrm{diag}\left(\mathbf{\Lambda}_{k}^{H}\left(\mathbf{\Lambda}_{k}\mathbf{P}_{k}^{\left(t\right)}\mathbf{\Lambda}_{k}^{H}+\mathbf{F}_{k}^{\left(t-1\right)}+\sigma^{2}\mathbf{I}_{r}\right)^{-1}\mathbf{\Lambda}_{k}\right)=h\mathbf{I}_{N_{t}}\label{eq:iterative_equation_solve_bisection}
\end{equation}
Introducing an auxiliary parameter $h$, one obtains an iterative
equation of $\mathbf{P}_{k}$. Without loss of generality, we assume
that the solution of (\ref{eq:iterative_equation_solve_eq}) belongs
to the feasible action set for given $\mathbf{P}_{-k}$. Otherwise,
similar analysis can applied for $\mathbf{P}_{k}$ but restricted
on a boundary given by Prop. \ref{prop:NE_EE_MIMO_continuous}. For
the sake of simplicity, we omit the discussion here and restrict ourselves
to the situation where the BR is strictly included in the interior
of the feasible action set. Therefore for all $i\in\left[N_{t}\right]$,
there exists $p_{ki}^{\star}$ such that individual utility function
$u_{k}\left(\mathbf{P}_{k},\mathbf{P}_{-k}\right)$ is an increasing
function in $\left(0,p_{ki}^{\star}\right)$ and a decreasing function
in $\left(p_{ki}^{\star},+\infty\right)$ with respect to $p_{ki}$,
where $p_{ki}^{\star}$ is the $i$-th component of user $k$'s BR
for given $\mathbf{P}_{-k}$. Then $u_{k}$ is also an increasing
function in $\left(0,U\left(\mathbf{P}_{-k}\right)\right)$ and a
decreasing function in $\left(U\left(\mathbf{P}_{-k}\right),+\infty\right)$
w.r.t. parameter $h$. In other words, to find $\mathbf{P}_{k}=\textrm{BR}\left(\mathbf{P}_{-k}\right)$,
it is sufficient to find $U\left(\mathbf{P}_{-k}\right)$ by a bisection
search due to the special monotonicity of the utility function.

However, it is worth mentioning that it is still difficult to directly
find the solution of iterative equation (\ref{eq:iterative_equation_solve_bisection}).
because this solution is actually implicitly given. We would like
to further simplify (\ref{eq:iterative_equation_solve_bisection})
to facilitate the calculation of BR or NE. To start with, we assume
that $N_{t}=N_{r}$. Firstly, we remove the diagonal operator of LHS
of (\ref{eq:iterative_equation_solve_bisection}). Therefore we have:
\begin{equation}
\mathbf{P}_{k}^{\left(t\right)}=\frac{1}{h}\mathbf{I}_{N_{t}}-\mathbf{\Lambda}_{k}^{-1}\left(\mathbf{F}_{k}^{\left(t-1\right)}+\sigma^{2}\mathbf{I}_{N_{r}}\right)\mathbf{\Lambda}_{k}^{-1}\label{eq:best_response_appr_Nt_eq_Nr}
\end{equation}
If $N_{t}>N_{r}$ or $N_{t}<N_{r}$ then $\mathbf{\Lambda}_{k}$ is
not directly invertible, then we should consider the pseudo-inverse
matrix of $\mathbf{\Lambda}_{k}$. Without loss of generality, we assume
that $N_{t}>N_{r}$, denoting the right pseudo-inverse of $\mathbf{\Lambda}_{k}$
as $\mathbf{\Lambda}_{k}^{\dagger}$ then one has $\mathbf{\Lambda}_{k}\mathbf{\Lambda}_{k}^{\dagger}=\mathbf{I}_{N_{r}}$
and $\left(\mathbf{\Lambda}_{k}^{\dagger}\right)^{H}\mathbf{\Lambda}_{k}^{H}=\mathbf{I}_{N_{r}}$.
Similarly, one has:
\begin{align}
\mathbf{\Lambda}_{k}^{H}\left(\mathbf{\Lambda}_{k}\mathbf{P}_{k}^{\left(t\right)}\mathbf{\Lambda}_{k}^{H}+\mathbf{F}_{k}^{\left(t-1\right)}+\sigma^{2}\mathbf{I}_{r}\right)^{-1}\mathbf{\Lambda}_{k} & =h\mathbf{I}_{N_{t}}\notag\\
\left(\mathbf{\Lambda}_{k}\mathbf{P}_{k}^{\left(t\right)}\mathbf{\Lambda}_{k}^{H}+\mathbf{F}_{k}^{\left(t-1\right)}+\sigma^{2}\mathbf{I}_{r}\right)^{-1} & =h\left(\mathbf{\Lambda}_{k}^{\dagger}\right)^{H}\mathbf{\Lambda}_{k}^{\dagger}\label{eq:appr_BR_step1}
\end{align}
However, it is generally impossible to have $\mathbf{\Lambda}_{k}^{\dagger}\mathbf{\Lambda}_{k}=\mathbf{I}_{N_{t}}$.
Thus the equality does not always holds when we multiply $\mathbf{\Lambda}_{k}^{\dagger}$
on left and $\left(\mathbf{\Lambda}_{k}^{\dagger}\right)^{H}$ on
the right on both sides of the equation. Nevertheless, this operation
will yield a linear approximation of the BR dynamics:
\begin{equation}
\widehat{\mathbf{P}}_{k}^{\left(t\right)}=  \frac{\mathbf{\Lambda}_{k}^{\dagger}\left[\left(\mathbf{\Lambda}_{k}^{\dagger}\right)^{H}\mathbf{\Lambda}_{k}^{\dagger}\right]^{-1}\left(\mathbf{\Lambda}_{k}^{\dagger}\right)^{H}}{h}-\mathbf{\Lambda}_{k}^{\dagger}\left(\mathbf{F}_{k}^{\left(t-1\right)}+\sigma^{2}\mathbf{I}_{N_{r}}\right)\left(\mathbf{\Lambda}_{k}^{\dagger}\right)^{H}\label{eq:epsilon_appr_BR_Nt_geq_Nr}
\end{equation}
Similarly, if $N_{t}<N_{r}$ we can obtain exactly the same iterative
equation as (\ref{eq:epsilon_appr_BR_Nt_geq_Nr}). This type of dynamics
belongs to the so-called $\varepsilon$-approximate best response
which generally leads to the $\varepsilon$-approximate Nash Equilibrium
defined as:
\begin{definition}
For game $\mathcal{G}=\left(\mathcal{K},\left(\mathcal{P}_{k}\right)_{k\in\mathcal{K}},\left(u_{k}\right)_{k\in\mathcal{K}}\right)$,
an action profile $\boldsymbol{p}=\left(\boldsymbol{p}_{k},\boldsymbol{p}_{-k}\right)$
is called an $\varepsilon$-approximate Nash Equilibrium if for $\forall k\in\mathcal{K}$
and $\boldsymbol{p}'=\left(\boldsymbol{p}'_{k},\boldsymbol{p}_{-k}\right)$
for $\varepsilon\geq0$:
\begin{equation}
u_{k}\left(\boldsymbol{p}_{k},\boldsymbol{p}_{-k}\right)-u_{k}\left(\boldsymbol{p}'_{k},\boldsymbol{p}_{-k}\right)\geq-\varepsilon\label{eq:defintion_approximate_NE}
\end{equation}
\end{definition}
Obviously $\varepsilon$-approximate Nash Equilibrium is actually
an extension of the concept of Nash Equilibrium. Notice that when
$\varepsilon=0$ then we are exactly back to the definition of Nash
Equilibrium. If one deploys (\ref{eq:epsilon_appr_BR_Nt_geq_Nr})
as the BR dynamics to compute NE according to alg. \ref{alg:Bisection_Algorithm_solve_equation},
one may only result in $\varepsilon$-approximate Nash Equilibrium
of the game. To this end, we obtain a sub-optimal algorithm summarized
in alg. \ref{alg:Bisection_Algorithm_solve_equation} by using the
iterative equation deduced in (\ref{eq:epsilon_appr_BR_Nt_geq_Nr})
instead of using (\ref{eq:iterative_equation_solve_eq}).
\begin{algorithm}[ht]
\begin{algorithmic}

\STATE\ Initialization: $\mathbf{P}_{k}^{\left(0\right)}=\frac{1}{N_{t}}\mathbf{I}_{N_{t}},\forall k$.
choose $T$, $\epsilon_{1}$ and $\epsilon_{2}$

\STATE\ $\mathbf{For}$ $t=1$ to $T$, $\mathbf{do}$

\STATE\ \ \ $\mathbf{For}$ $k=1$ to $K$, $\mathbf{do}$

\STATE\ \ \ \ \ Initialization: $\underline{h}=0$ and $\overline{h}=h_{max}$

\STATE\ \ \ \ \ $\mathbf{Repeat}$ Until $\overline{h}-\underline{h}\leq\epsilon_{1}$

\STATE\ \ \ \ \ \ \  $h_{L}=\max\left(0,h_{M}-\frac{\epsilon_{1}}{2}\right)$
and $h_{M}=\frac{\underline{h}+\overline{h}}{2}$

\STATE\ \ \ \ \ \ \  $h_{R}=\min\left(h_{max},h_{M}+\frac{\epsilon_{1}}{2}\right)$

\STATE\ \ \ \ \ \ \ Compute $\mathbf{P}_{k}\left(h_{i}\right)$
using (\ref{eq:epsilon_appr_BR_Nt_geq_Nr}), $i\in\left\{ L,M,R\right\} $

\STATE\ \ \ \ \ \ \  $U_{i}=u_{k}\left(\mathbf{P}_{k}\text{\ensuremath{\left(h_{i}\right)}},\mathbf{P}_{-k}^{\left(t-1\right)}\right)$,
$i\in\left\{ L,M,R\right\} $

\STATE\ \ \ \ \ \ \ $\mathbf{If}$ $U_{L}<U_{M}<U_{R}$

\STATE\ \ \ \ \ \ \ \ \ \ \ $\underline{h}=h_{L}$

\STATE\ \ \ \ \ \ \ \ \ \ \ $\mathbf{Else}$ $\mathbf{If}$ $U_{L}>U_{M}>U_{R}$

\STATE\ \ \ \ \ \ \ \ \ \ \ \ \ \ \  $\overline{h}=h_{R}$

\STATE\ \ \ \ \ \ \ \ \ \ \  $\mathbf{End}$ $\mathbf{If}$

\STATE\ \ \ \ \ \ \ $\mathbf{Else}$ 

\STATE\ \ \ \ \ \ \ \ \ \ \ $\underline{h}=h_{L}$ and
$\overline{h}=h_{R}$

\STATE\ \ \ \ \ \ \ $\mathbf{End}$ $\mathbf{If}$

\STATE\ \ \ \ \ \ \ Compute $\mathbf{P}_{k}^{\left(t\right)}$
by (\ref{eq:epsilon_appr_BR_Nt_geq_Nr}) with $h=h_{M}$

\STATE\ \ \ $\mathbf{End}$ $\mathbf{For}$

\STATE\ \ \ $\mathbf{If}$ $\sum_{k}\left\Vert \mathbf{P}_{k}^{\left(t\right)}-\mathbf{P}_{k}^{\left(t-1\right)}\right\Vert <\epsilon_{2}$

\STATE\ \ \ \ \ \ \ $\mathbf{Break}$

\STATE\ \ \ $\mathbf{End}$ $\mathbf{If}$

\STATE\ $\mathbf{End}$ $\mathbf{For}$

\STATE\ Output: $\mathbf{P}_{k}^{\textrm{NE}}=\mathbf{P}_{k}^{\left(t\right)}$
for $\forall k$.

\end{algorithmic}

\caption{Bisection Search Algorithm for find the NE of MIMO-MAC EE game  \label{alg:Bisection_Algorithm_solve_equation}}
\end{algorithm}

\begin{remark}
Alg. \ref{alg:Bisection_Algorithm_solve_equation} actually works
for general utility function possessing the same property as $u_{k}$.
Moreover, this algorithm should be slightly faster than general bisection
search. The reason is once by coincidence
that the case neither $U_{L}>U_{M}>U_{R}$ nor $U_{L}<U_{M}<U_{R}$
occurs, we are surely to be very close to the stationary point of
function. Otherwise
we are in the monotonic region of the function, then this algorithm
works as regular bisection search algorithm. In the worst case, this
algorithm should have same complexity as the general bisection algorithm.
Finally, alg. \ref{alg:Bisection_Algorithm_solve_equation} merely
requires the value of utility function instead of derivative of the
utility function. Notice that alg. 2 is actually an off-line
learning algorithm. Therefore a online-learning-version of alg.
2 by combing it with some deep learning techniques could be
an important extension of this paper.
\end{remark}

\section{Numeric Results\label{sec:Numeric_Results}}
The goal of this part is to show the performance of the proposed algorithms.
Notice if $N_{t}=N_{r}$, (\ref{eq:epsilon_appr_BR_Nt_geq_Nr}) degenerates
to (\ref{eq:best_response_appr_Nt_eq_Nr}) which conserves the optimality
of best response. For this situation, we choose $N_{t}=N_{r}=2$ with
$K=2$ users. A sufficient large power budget is chosen so that
the BR is included in the feasible action set $\overline{P}_{k}=10mW$
for $\forall k\in\left\{ 1,2\right\} $ and the circuit power is $P_{c}=1mW$.
The error tolerance for alg. \ref{alg:Bisection_Algorithm_solve_equation}
is $\epsilon_{1}=\epsilon_{2}=0.001$.
\begin{figure}[ht]
\begin{centering}
\includegraphics[scale=0.6]{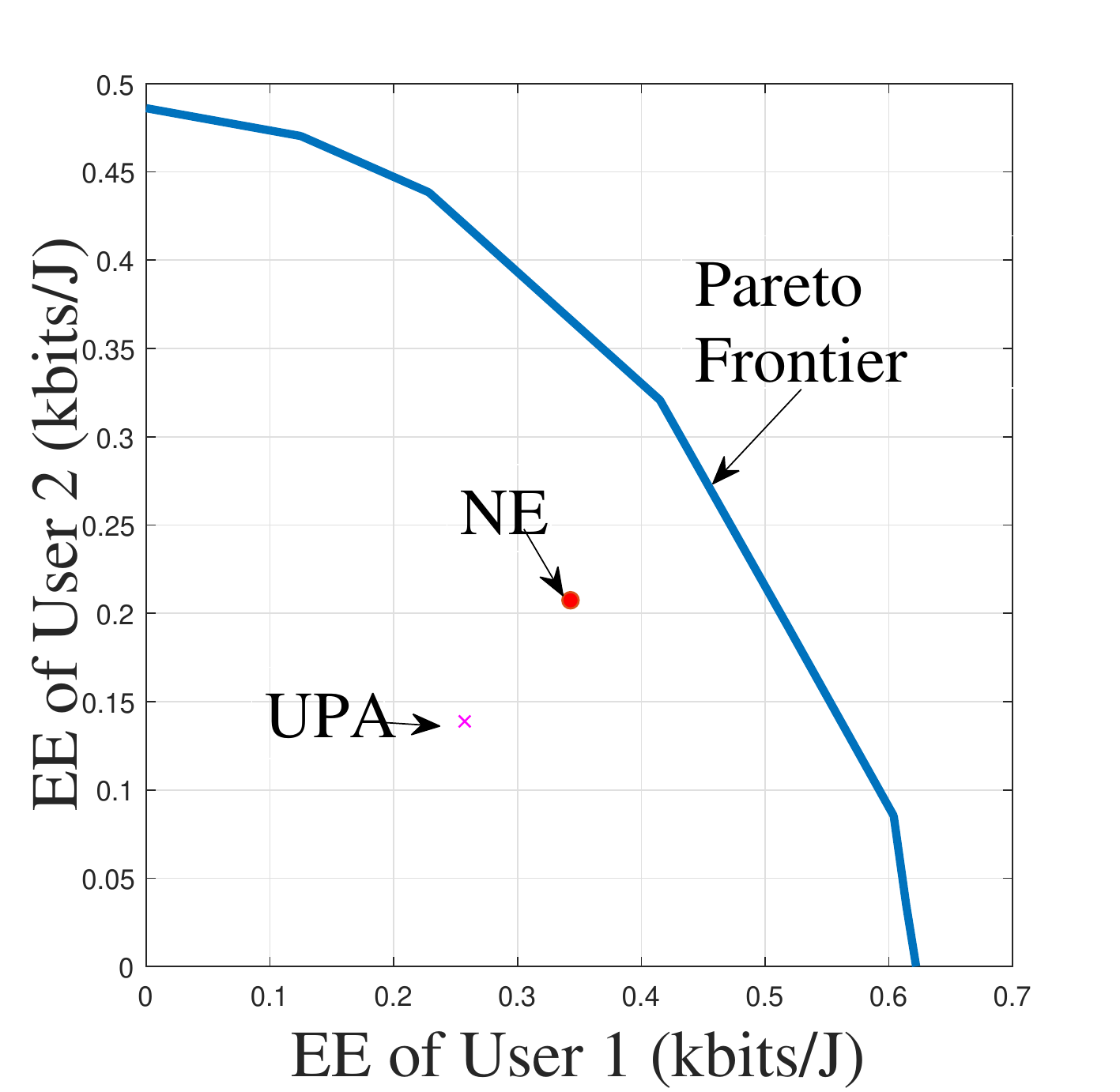}
\par\end{centering}
\caption{Energy Efficiency under NE and uniform power allocation with $N_{t}=N_{r}=2$
for $2$-user situation. Policy found by our algorithms outperforms
 UPA policy.\label{fig:Nt_2_Nr_2}}
\end{figure}

\begin{figure}[ht]
\begin{centering}
\includegraphics[scale=0.6]{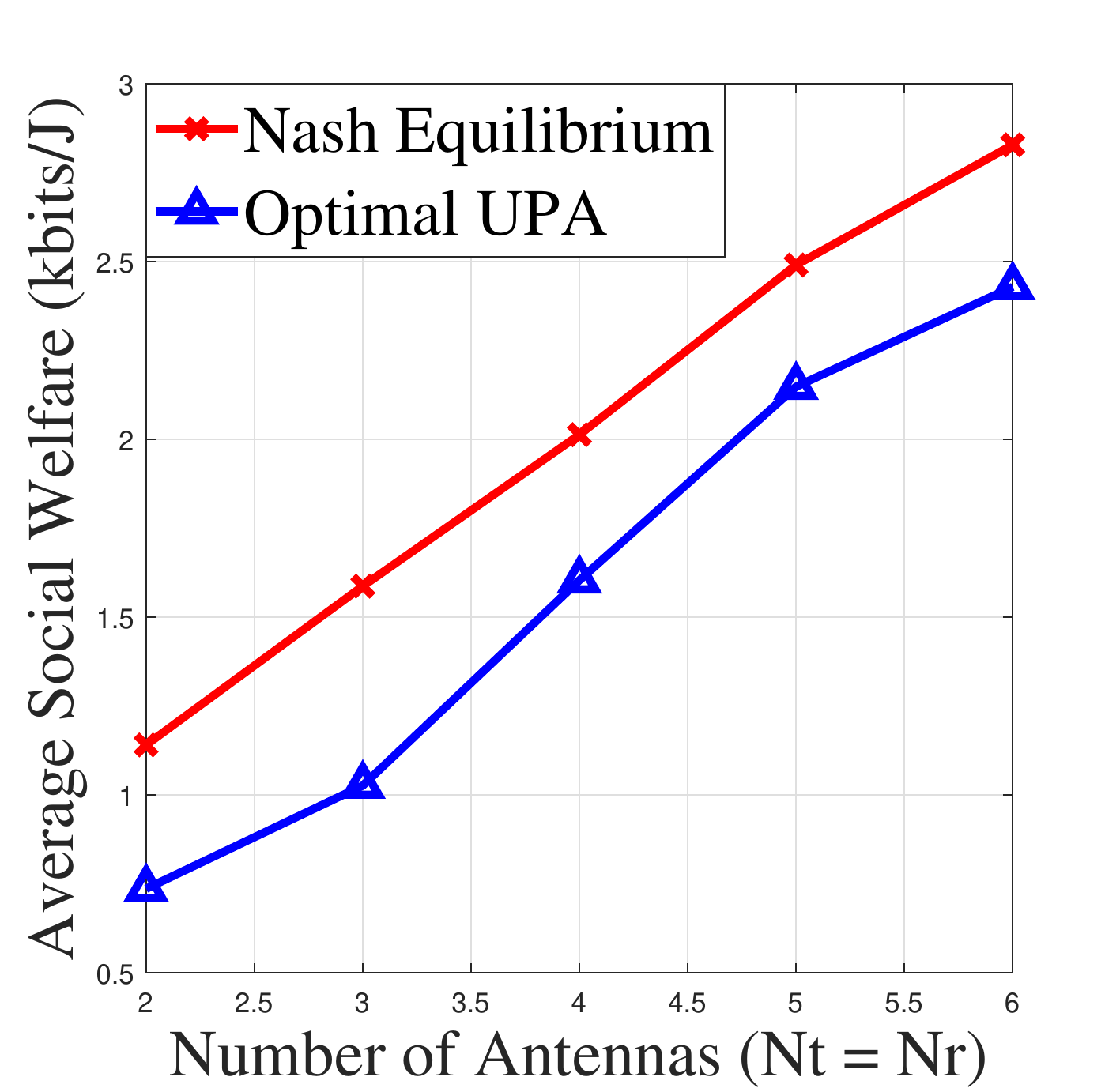}
\par\end{centering}
\caption{Average social welfare under NE and UPA as function of number of antennas
($N_{t}=N_{r}$) with $\overline{P}_{k}=10mW$ for $2$-user situation.\label{fig:eq_antenna_vary}}
\end{figure}

In Fig. \ref{fig:Nt_2_Nr_2}, the achievable utility region, the average
performance under NE found by alg. \ref{alg:Bisection_Algorithm_solve_equation}
and the averaged performance achieved by uniform power allocation
(UPA) are depicted. All results are averaged over $1000$ randomly
generated channel samples. It is observed that the performance achieved
by deploying UPA is Pareto-dominated by
NE which can be found by alg. \ref{alg:Bisection_Algorithm_solve_equation}.
Furthermore, the NE found by alg. \ref{alg:Bisection_Algorithm_solve_equation}
is closed to the Pareto frontier achieved by some centralized algorithms
which suggest the efficiency using alg. \ref{alg:Bisection_Algorithm_solve_equation}
is higher than UPA.

\begin{figure}[ht]
\begin{centering}
\includegraphics[scale=0.6]{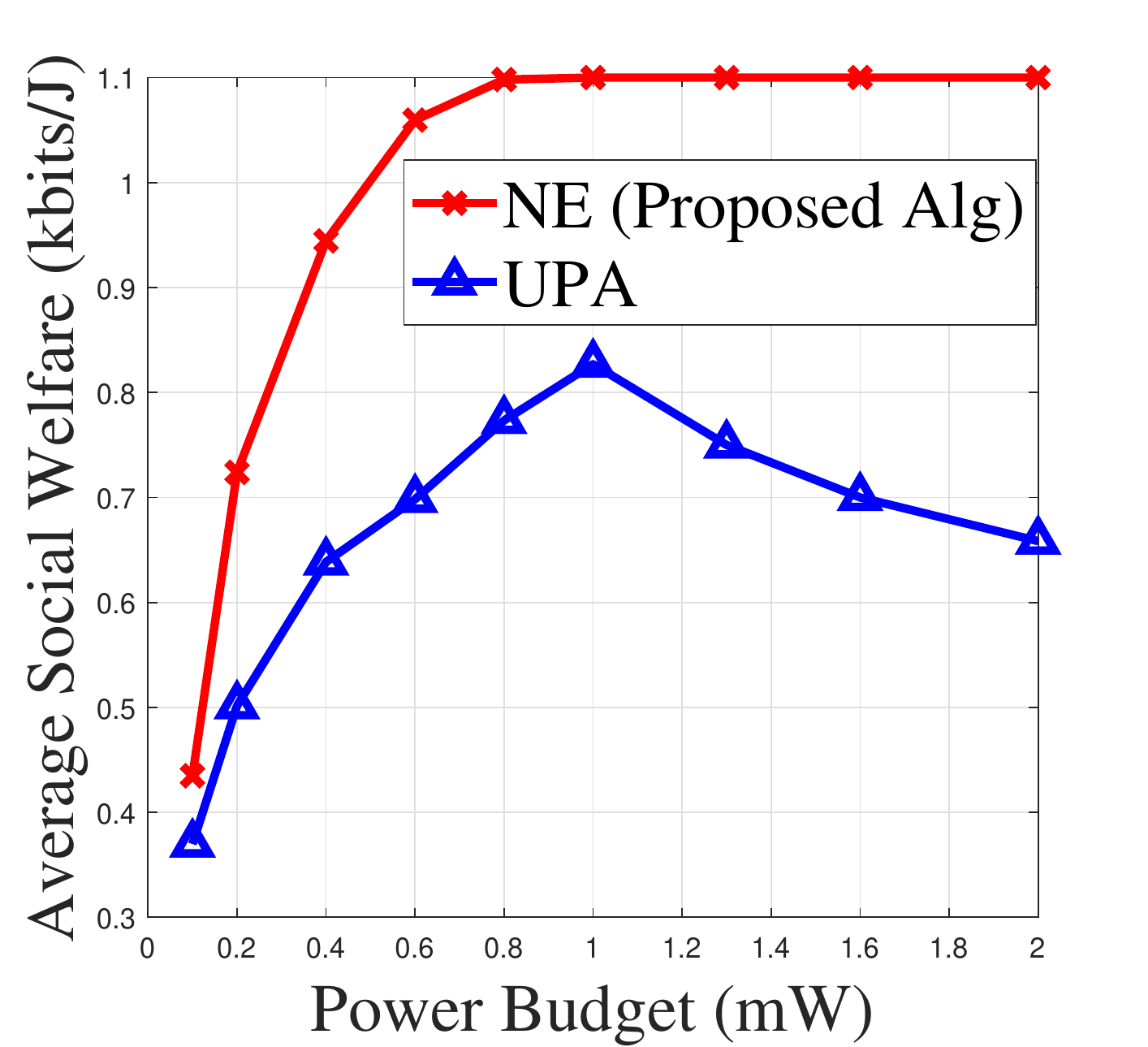}
\par\end{centering}
\caption{Performance under NE and UPA as function of the power budget of user
with $N_{t}=N_{r}=2$ for $2$-user situation. There are two different
regions: one corresponds to Prop. \ref{prop:NE_EE_MIMO_continuous}.
In the region uncovered by Prop. \ref{prop:NE_EE_MIMO_continuous}, proposed algorithm still dominates UPA. \label{fig:eq_antenna_power_vary}}
\end{figure}

\begin{figure}[ht]
\begin{centering}
\includegraphics[scale=0.6]{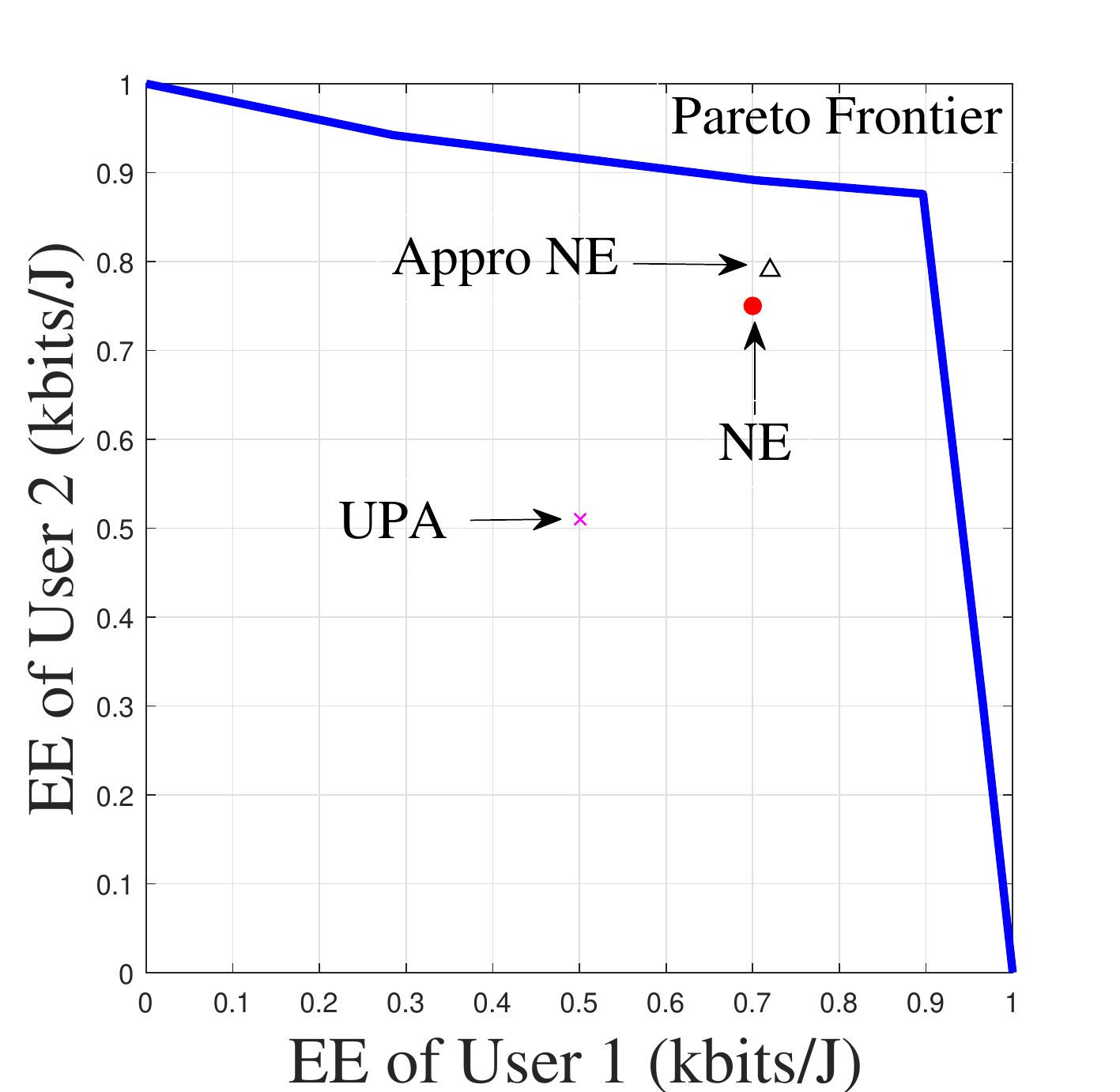}
\par\end{centering}
\caption{Performance achieved by alg. \ref{alg:basic_algorithm_for_NE} (NE)
and alg. \ref{alg:Bisection_Algorithm_solve_equation} (Approximate
NE) and UPA with $N_{t}=2$ and $N_{r}=4$ for $2$-user situation.
Policy found by alg. \ref{alg:Bisection_Algorithm_solve_equation}
is very near to the exact NE and Pareto-dominates it. Moreover, two
policies found by proposed algorithms both outperform UPA.\label{fig:Nt_2_Nr_4}}
\end{figure}

Moreover, define the social welfare for a given action profile as
$w\left(\boldsymbol{p}\right)=\sum_{k\in\mathcal{K}}u_{k}\left(\boldsymbol{p}_{k},\boldsymbol{p}_{-k}\right)$. 
Then the average social welfare as function of number of number of
antennas (still we keep $N_{t}=N_{r}$) and the power budget 
in Fig. \ref{fig:eq_antenna_vary} and Fig. \ref{fig:eq_antenna_power_vary}
respectively. For Fig. \ref{fig:eq_antenna_vary}, the averaged social
welfare of both UPA policy and our proposed algorithm is increased
quasi-linearly as the number of antennas grows. However our proposed
algorithm always outperforms the optimal UPA policy which is allowed
to tune the power but always equally shared among each transmit antenna. In Fig. \ref{fig:eq_antenna_power_vary}, we would like to show the
influence of user's power budget. There are two different regions
for social welfare. In the first region where the power budget is
sufficiently large, the NE found by our proposed algorithm is independent
of the power budget while the performance of UPA is  decreasing with respect to
the increase of the power budget. In the second region where the
power budget is relatively small, Using proposed algorithm, it is
not sure to converge to the NE of the game because Prop. \ref{prop:NE_EE_MIMO_continuous}
is no more valid in this region. Nevertheless, the performance achieved
by our algorithm is still better than UPA which prove the superiority
of our algorithm.

Then a more probable situation is considered where $N_{t}<N_{r}$
meaning that the number of antennas in user terminal is less than
the one in base station. The discussion in Sec. \ref{sec:Algorithm_for_NE}
shows that the proposed suboptimal algorithm is actually suboptimal
due to the usage of $\varepsilon$-approximate best response. For
numeric demonstration, we choose $N_{t}=2<N_{r}=4$. The performance
of alg. \ref{alg:Bisection_Algorithm_solve_equation} is illustrated
in Fig. \ref{fig:Nt_2_Nr_4}. The sub-optimality is clearly demonstrated
in this figure. However, the resulted policy actually Pareto-dominates
the exact NE found by alg. \ref{alg:basic_algorithm_for_NE} and the
dispersion is relatively small in terms of average performance. This
remark entails that even the policy found by alg. \ref{alg:Bisection_Algorithm_solve_equation}
is not the NE of the game in its sub-optimal region however its performance
does slightly outperforms the exact NE. Moreover the proposed algorithm
is easy to implement for using explicit iterative equation even if
it is approximated.

\section{Conclusions\label{sec:Conclusions}}
In this paper, a game where the individual utility function is the
energy efficiency in a MIMO multiple access channel system is considered.
The existence and the uniqueness of Nash Equilibrium is proved and
an exact algorithm and a suboptimal algorithm is proposed to find
the NE of this game. Simulation results show that if the the number
of transmit antennas and the number of receiving antennas is the same,
performance under NE found by proposed algorithms is always better
than uniform power allocation policy for both inside or outside the range covered by
the main proposition of the paper. When the condition for antennas
is not met, our proposed algorithm actually deploys an $\varepsilon$-approximate
best response which might leads to an $\varepsilon$-approximate Nash Equilibrium.
Quiet surprisingly the approximate NE found by our sub-optimal algorithm
slightly Pareto-dominates the exact NE of the game. This observation
shows that the performance of proposed algorithm is acceptable while
it is relatively easy to implement. Other techniques such as pricing
might be useful to improve the efficiency of the overall system. The
situation where each user is allowed to freely choose its covariance
matrix merely constrained to the maximum power is the natural extension
of this paper. Moreover, the discussion over the effect of successive
interference cancellation and multiple carrier seems to be complicated
and serve as the challenge of the future works.

\end{document}